\documentstyle[pre,twocolumn,aps,epsfig,floats]{revtex}

\begin{document}
\draft

\title{ Daisy models: Semi-Poisson statistics and beyond}
\author{H. Hern\'andez-Salda\~na$^{\dagger}$, 
J. Flores$^{\dagger \ddagger}$  and T.H. Seligman}
\address {Centro Internacional de Ciencias}
\address{and}
\address {Centro de Ciencias F\'{\i}sicas, University of Mexico (UNAM)}
\address{Ciudad Universitaria, Chamilpa, Cuernavaca, M\'exico}
\maketitle 

\begin{abstract}

Semi-Poisson statistics are shown to be obtained  by removing 
every other number from a random sequence. Retaining every 
$( r+1 )$th level we obtain a family of sequences which we call daisy models.
Their statistical properties coincide 
with those of Bogomolny's nearest-neighbour interaction Coulomb gas if
the inverse temperature coincides with the integer $r$. In particular
the case $ r = 2 $
reproduces closely the statistics of quasi-optimal solutions of 
the traveling salesman problem.
\end{abstract}

The transition from order to chaos in a classical system is generically 
reflected by a 
transition in spectral statistics for the corresponding quantum system 
from Poisson to GOE statistics. Such a transition was first considered by 
Porter and Rosenzweig \cite{port} in a model that features a parameter
that depresses 
the off-diagonal elements of a GOE until they are zero and we have 
Poisson statistics. This model is amenable to analytical treatment \cite{Leyv}, 
but does not reflect the properties of dynamical systems very well. 
Band matrices were introduced later \cite{selig} and have been quite 
successful in the description of many situations \cite{izra}.
A semi-classical ansatz by Berry and Robnik \cite{berry} was shown to work very 
well if applied to the long-range behaviour of the two-point function 
\cite{selig2}.

More recently a different kind of transitional behaviour was discovered 
near delocalization transitions \cite{shkl} and in 
pseudo-integrable systems \cite{bogo,PhD}; it is
commonly known as semi-Poisson statistics. Bogomolny, Gerland and Schmit
developed a 
level dynamics for this type of spectra by limiting the usual $1-D$ 
Coulomb gas model 
to nearest-neighbour interactions and considering an inverse temperature 
$ \beta = 1 $ \cite{bogo}. They also pointed out that the 
nearest-neighbour spacing 
distribution
can be reproduced by an interpolation procedure in a Poisson spectrum 
\cite{bogo,PhD}.

 The first purpose of the present note is to show that the statistics of a 
 semi-Poisson spectrum are reproduced exactly by a model where every 
 other level is dropped from a Poisson spectrum. The idea for such a model 
derives from well-known results that relate the superposition of two GOEs 
\begin{enumerate}
\item[$ ^{\dagger} $]{ Permanent address: Instituto de F\'{\i}sica, 
University of Mexico (UNAM).
Apdo. Postal 20-364, 01000 M\'exico. D.F. M\'exico}

\item[$ ^{\ddagger} $] e-mail: jfv@servidor.unam.mx 
\end{enumerate}
 to a GUE or one GOE to a GSE by the same procedure. We shall term such 
 models {\it daisy} models. As in the above cases a dynamical 
 link is not established. Rather we calculate properties of the 
 spectra and find that they coincide. This is certainly no dynamical 
 explanation of the properties of the physical systems that display 
 semi-Poisson statistics, but neither is any of the level-dynamics models.
 The main advantage is that it is a  simple model for which it is very 
 easy to compute any statistic.

 It is further interesting to note 
 that this procedure does not only yield one new type of statistical 
 spectra but rather an entire 
family of daisy models of rank $ r $ where $ r $ is the number of 
levels dropped 
between each retained level. Their statistical properties can be easily 
 calculated for any $ r $. As we shall see, the dependence on rank is exactly 
 the same as on 
 the inverse temperature $\beta$ in the nearest-neighbour interaction 
Coulomb gas. 
 Thus we find that all integer values of
 this parameter correspond to a daisy model. 

 No link of such statistical spectra to 
quantized dynamical systems is known for $r > 1$, but the 
case $ r = 2 $ displays 
surprising similarity to the statistical distribution of distances between 
cities along a 
 quasi-optimal path of the traveling salesman problem \cite{mendez} . It may be 
 worthwhile to note the relation of this problem to spin glasses \cite{spin} 
 though we shall not discuss this aspect. 

 The semi-Poisson spectra display a nearest-neighbour spacing 
 distribution
 $$
 P(s)=4s\exp(-2s)      \eqno (1)
 $$
 and a long-range behaviour of the two-point function defined by
 a number variance 
 $$
 \Sigma ^2 (L) = L/2.  \eqno (2)
 $$

 The nearest-neighbour interaction $1-D $ Coulomb gas model is defined
as follows:
 We have $N+2$ particles with positions $x_j$ in an interval of
 size $L$ with the interaction
 $$V(x_0,x _1,x _2,...,x_{N+1})= -\sum_{i}\log(x_i-x_{i-1})  \eqno (3)$$
 and  the condition $0= x_0 < x_1 < .... < x_N< x_{N+1} = L$.
 This model has the following $n$th-neighbour spacing
 distribution for inverse temperature $ \beta $
 $$P(n,s)=\frac {(\beta+1)^{(\beta+1)n}}{\Gamma([\beta+1]n)}s^{(\beta+1)n-1}
 \exp[-(\beta+1)s].  \eqno  (4)$$
 In particular for $\beta=1$ and $n=1$ we obtain the 
nearest-neighbour distribution
 for the semi-Poisson (eq. 1).
   
   For the daisy model of order $r$  as defined above we obtain the 
   $n$th-neighbour spacing distribution by rescaling 
the $rn$th-neighbour distribution
   of a random sequence as 
$$P_r(n,s)=\frac {(r+1)^{(r+1)n}}{\Gamma([r+1]n)}s^{(r+1)n-1}\exp[-(r+1)s]. 
\eqno (5)$$

 The same rescaling argument yields for the asymptotic behaviour with 
$L \gg 1$ of the
number variance for rank $r$ model, ${\Sigma_{r} ^2}(L) \sim L/(r+1)$. 
Using the expression (4.41) of ref. \cite{PhD} for the two-point function we 
obtain for the number variance

$$
{\Sigma_{r}^2}(L)=
L+\frac{2L}{r+1}\sum_{j=1}^r \frac{W_j}{(1-W_j)}  
+\frac{2}{{(r+1)}^2}\times   \eqno (6)
$$
$$\times\bigg(-\sum_{j=1}^r \frac{W_j}{(1-W_j)^2}
+\sum_{j=1}^r \frac{W_j}{(1-W_j)^2} \exp([W_j-1](r+1)L) \bigg)$$
where $W_j=\exp(\frac{2 \pi ij}{r+1})$ are the $r+1$ roots of unity.
All sums in this expression are real and the fist two can be summed as 
follows:

Consider the function $P(x)=\frac{x^{r+1}-1}{x-1}={\Pi_{j=1}^r}(x-W_j)$,
with $W_j$ as above, and its logarimic derivatives. 
We note that  $\bigg( \frac{d}{dx}\log(P(x)) \bigg)\bigg|_{x=1}=r/2=
\sum_{j=1}^r \frac{1}{1-W_j}$ and 
$-\bigg( \frac{d^2}{dx^2}\log(P(x)) \bigg)\bigg|_{x=1}=
\frac {-r^2+4r}{12}=\sum_{j=1}^r \frac{1}{(1-W_j)^2}$.
Hence, it is possible to write for the first sum
$\sum_{j=1}^r \frac{W_j}{1-W_j}=\sum_{j=1}^r \frac{W_j-1}{1-W_j} + 
\sum_{j=1}^r \frac{1}{1-W_j}=-r/2$, and for the second one 
$$
\sum_{j=1}^r \frac{W_j}{(1-W_j)^2}=\sum_{j=1}^r \frac{W_j-1}{(1-W_j)^2}
+\sum_{j=1}^r \frac{1}{(1-W_j)^2} \eqno(7)$$
$$= -\frac{r}{2}+\frac {-r^2+4r}{12}=
\frac {-r(r+2)}{12}.
$$ 

We thus obtain

$$
{\Sigma_{r}^2}(L)=\frac{L}{r+1}+\frac{r(r+2)}{6{(r+1)}^2}  \eqno(8)$$
$$
 +\frac{2}{{(r+1)}^2}
\sum_{j=1}^r \frac{W_j}{(1-W_j)^2} \exp([W_j-1](r+1)L),
$$
which corroborates the asymptotic behaviour for $ L \gg 1$.
In particular, for $r=2$ we have 
$$
{\Sigma_{2}^2}(L)=\frac{L}{3}+\frac{4}{27}\bigg [ 1-\cos(\frac{3 \sqrt{3}}{2}L)
\exp(-\frac{9}{2}L)\bigg ], \eqno (9)
$$
which is depicted by the solid line in fig 1.

\begin{figure}[htb]
\begin{center}
 \leavevmode
{\psfig{file=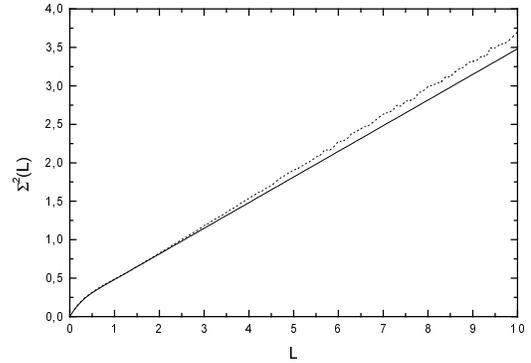,height=7.8cm,angle=270}}
\end{center}
\caption{Fig. 1: The number variance $\Sigma^2$ for the traveling salesman 
problem (dashed line) compared to the one of the rank $2$ daisy model 
(full line).} \label {fig:1}
\end{figure}

 If we compare equations (4) and (5) we find that they coincide for 
$ \beta = r $. Thus, we link the inverse temperature in the Coulomb gas
model to
the number of discarded elements in a rank $ r $ daisy model starting from 
a random sequence. The fact that semi-Poisson statistics can be 
interpreted in this simple way is of interest 
because the properties 
of daisy models over random sequences are easy to calculate
and because it may shed some light into the  $1-D $ Coulomb gas dynamics.

 On the other hand, we may ask if models of rank larger than $1$ are 
relevant. Some of us pointed out recently \cite{mendez} that the 
statistical distribution of distances between
cities along a quasi-optimal path of the traveling salesman problem
displays characteristic features that can be analyzed with the tools of 
spectral statistics, but cannot be understood in terms of any of the 
usual random matrix models. In particular, the long-range behaviour 
of the number variance 
$\Sigma^2$ and the 
correlation coefficient between adjacent spacings seem quite incompatible 
with band matrices or Porter-Rosenzweig type models \cite{mendez}. We shall
therefore investigate whether we obtain a better agreement with a daisy 
model.

The data for the traveling salesman problem 
are obtained for an ensemble of 500 maps  of 500 cities using 
simulated anneling. 
We first compared with the pseudo-Poisson model and found no agreement. 
But the slope
of the number variance suggests that we should rather compare with the rank 
2 daisy model for a random sequence.

\begin{figure}[tbp]
{\hspace*{-.5cm}\psfig{figure=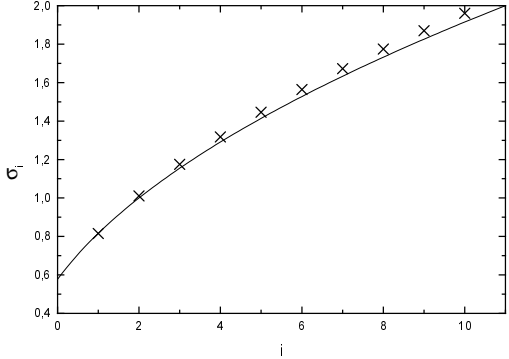,height=7.8cm,angle=-90}}
\vspace*{.12in}
\caption{Fig. 2: The width $\sigma_i$ of $i$th-neighbour spacing 
distribution of the
traveling salesman problem (crosses) compared to that of the rank $2$ 
daisy model (full line).} \label {fig:2}
\end{figure}

    In figure 1 we compare  the number variance with eq. (9) 
    for the rank 2 model and
    we find a remarkable agreement. If we fit the asymptotic of the number 
    variance ${\Sigma_{2}^2}(L)$ with a straight line  for the interval 
$2.0 \leq L \leq 9.0$ we 
obtain a slope of  $0.358\pm 0.001$
near to the value $ 1/3 $. Furthermore, the $i$th-neighbour  
spacing width $\sigma_i$ for the rank 2 model is equal to 
a rescaled $ 2i $th-neighbour 
distribution of a Poisson ensemble. In fig 2  we compare
these widths with the ones obtained for the traveling salesman problem in 
\cite{mendez} and find similar agreement. For the correlation coefficient in
the rank 2 model we expect zero, which is quite near to the 
value $ .036 \pm  .002$ of ref. \cite{mendez}. We also compared the 
properties of skewness and curtosis used commonly to detect properties of
the three and four point functions. 
The results for the rank 2 model were obtained numerically 
and the comparison is displayed in figs 3a and 3b.
Similarly the nearest-neighbour spacing distributions are compared in
fig. 4. 
The agreement is certainly comparable to the 
one obtained for 
pseudo-integrable systems with the semi-Poisson statistics.

\begin{figure}[tbp]
{\hspace*{-.5cm}\psfig{figure=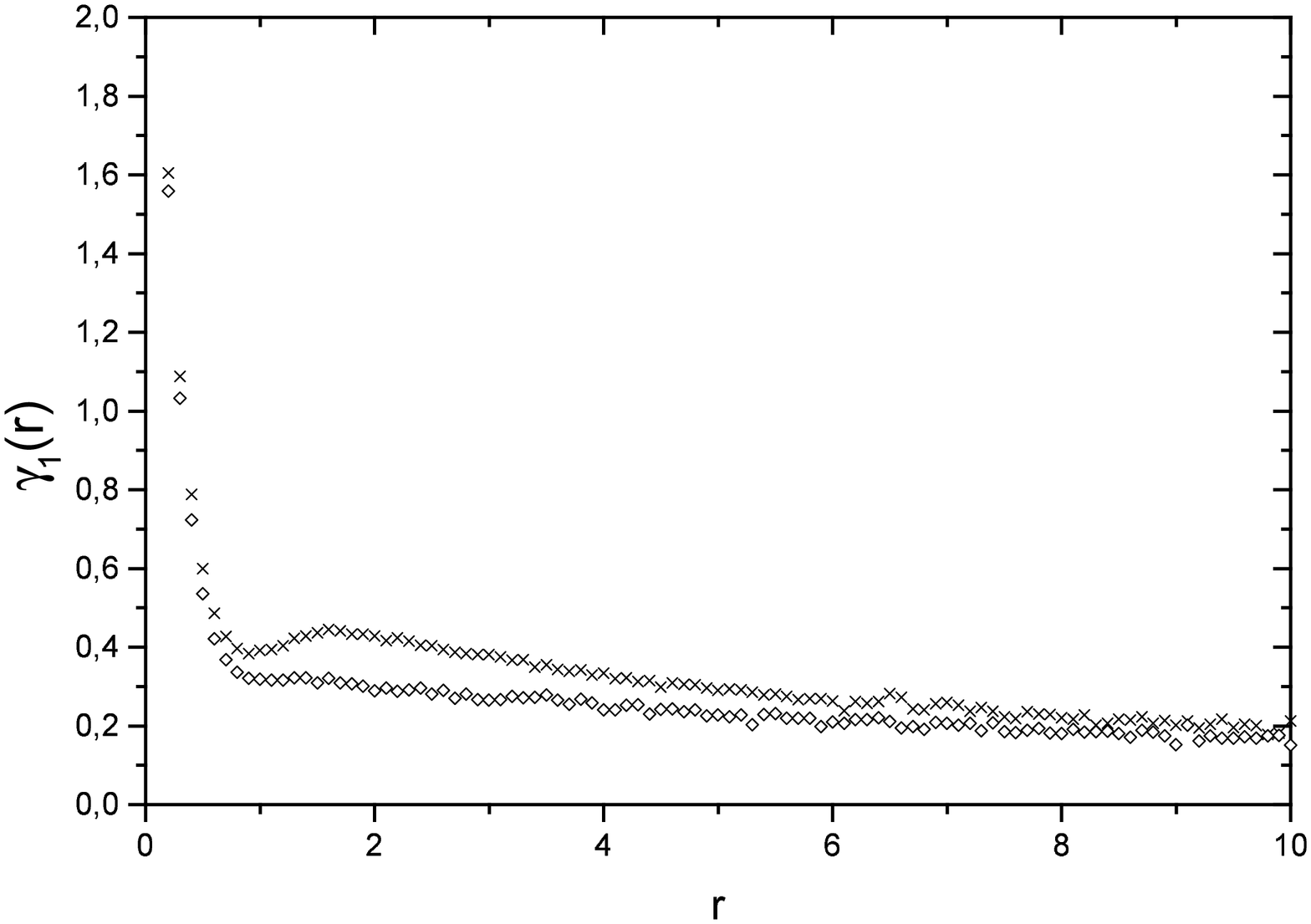,height=7.8cm,angle=270}}
{\hspace*{-.5cm}\psfig{figure=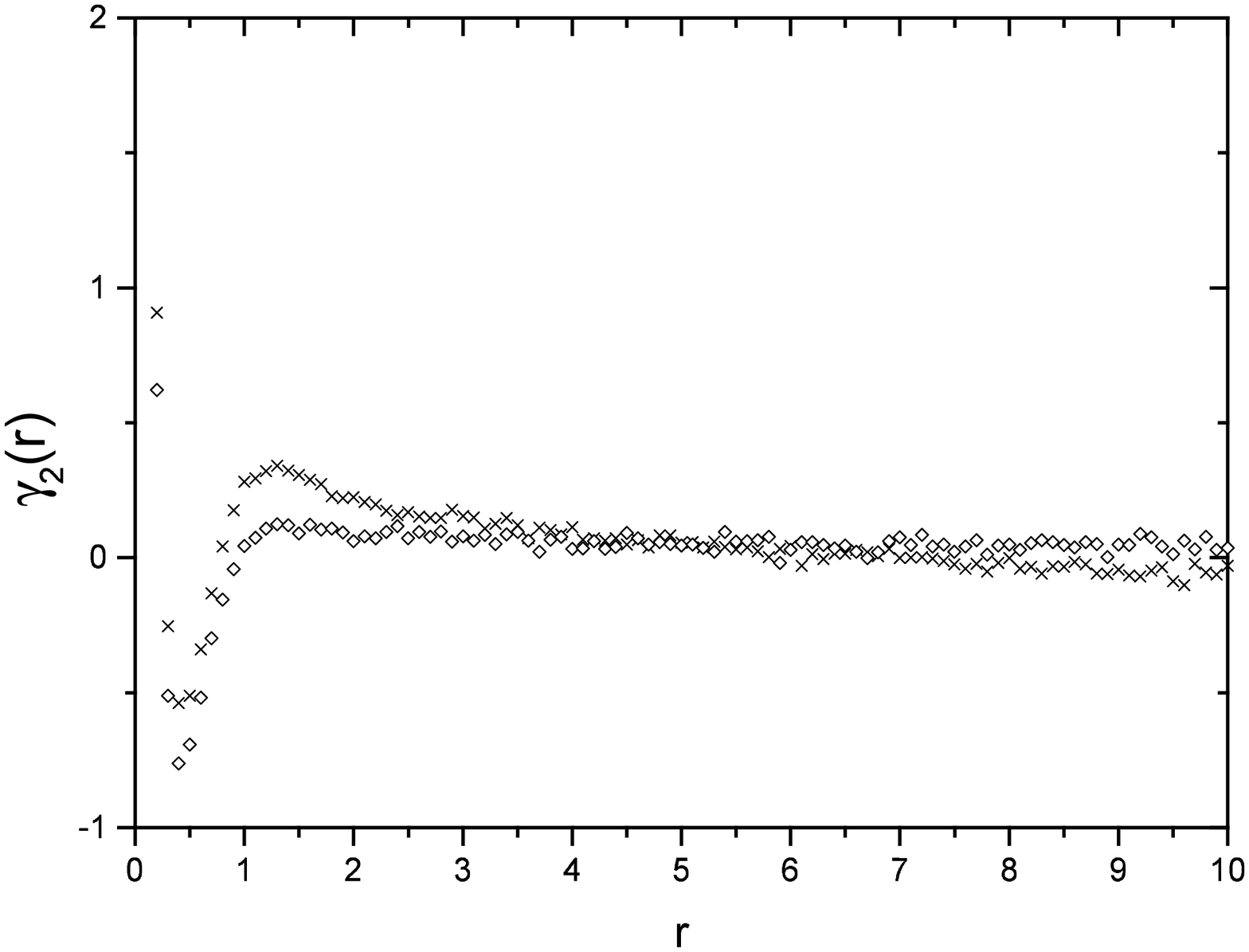,height=7.8cm,angle=270}}
\vspace*{.12in}
\caption{ The skewness (a) and the curtosis (b) of the 
traveling salesman problem
(crosses) compared to those of the  rank $2$ daisy model (diamonds).
} \label {fig:3b}
\end{figure}

\begin{figure}[tbp]
{\hspace*{-.5cm}\psfig{figure=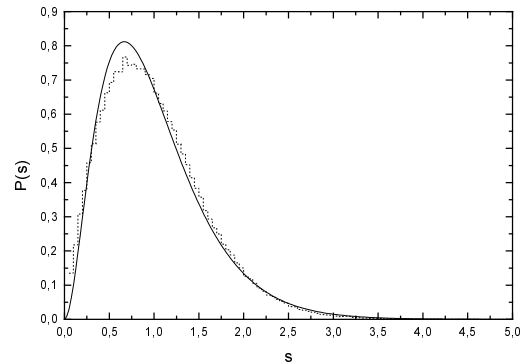,height=7.8cm,angle=270}}
\vspace*{.12in}
\caption{Fig. 4: Nearest-neighbour spacing  distribution for the traveling 
salesman problem
(dashed line) compared to that of the  rank $2$ daisy model (full lines)} 
\label {fig:4}
\end{figure}

     Summarizing, we have shown that semi-Poisson statistics can be obtained 
     from a very simple model without any dynamical implications.  
This model actually pertains to a family of models that seem to be relevant
in situations where the usual banded 
matrix models and the Porter Rosenzweig model are grossly inadequate.
We obtain this family by retaining every $(r+1)$th level of a random sequence.
A similar selection process could be performed  for the classical 
ensembles of Cartan (e.g. the GOE)\cite{cartan}. Whether this leads to 
useful results
beyond the two cases mentioned above is an open question.

{ \bf Acknowledgments}

We would  like to thank F. Leyvraz for helpful discussions. This project 
has been supported by DGEP, DGAPA IN-112998, UNAM and CONACYT 25192-E.

\end{document}